\documentstyle[12pt]{article}
\begin{document}
\makeatother
\renewcommand{\theequation}{\thesection.\arabic{equation}}
\newcommand{\am}{a \times m}
\newcommand{\ri}{RI }
\newcommand{\mathleftline}[1]{\hbox to0pt{\hss\hbox to12pt{\hbox
 to\hsize{$#1$\hfill}\hss}}}
\newcommand{\ot}{\frac{1}{2}}
\newcommand{\D}{& \displaystyle}  %instead of &
\newcommand{\di}{\displaystyle}   %at begin of each line

\date{October 1992}
\title
{Renormalization Group Analysis of Finite-Size Scaling
in the $\Phi^4_4$ Model\thanks{Supported by Fonds zur F\"orderung der
Wissenschaftlichen Forschung in \"Osterreich, project P7849.} }
\author
{\bf R. Kenna and C.B. Lang \\  \\
Institut f\"ur Theoretische Physik,\\
Universit\"at Graz, A-8010 Graz, AUSTRIA}
\maketitle
\begin{abstract}
A finite-size scaling theory for the $\phi^4_4$ model is derived using
renormalization group methods. Particular attention is paid to the
partition function zeroes, in terms of which all thermodynamic
observables can be expressed. While the leading scaling behaviour is
identical to that of mean field theory, there exist multiplicative
logarithmic corrections too. A non-perturbative test of these formulae
in the form of a high precision Monte Carlo analysis reveals good
quantitative agreement with the analytical predictions.
\end{abstract}

\newpage

%%%%%%%%%%%%%%%%%%%%%%%%%%%%%%%%%%%%%%%%%%%%%%%%%%%%%%%%%%%%%%%%%%%%%%%%
\section{Introduction}
%%%%%%%%%%%%%%%%%%%%%%%%%%%%%%%%%%%%%%%%%%%%%%%%%%%%%%%%%%%%%%%%%%%%%%%%
\setcounter{equation}{0}

Above one dimension, lattice $\phi^4$ theory is known to possess a
second order phase transition separating an ordered phase
from  a disordered one. The continuum parameterization of the field
theory is defined at this phase transition. There exist rigorous proofs
regarding the trivial (Gaussian) and interactive nature of the
continuum theory in $d>4$ and $d<4$ dimensions respectively
\cite{triviality}. Although a rigorous proof is still lacking, it is
believed that in $d=4$ the theory is also trivial. Nonetheless, the
theory may be useful as one with an effective interaction --- valid
below some momentum cutoff $\Lambda$. In this context, triviality of the
theory means that the (leading) critical exponents at the phase
transition are identical to those of the Gaussian model which describes
free bosons. The renormalization group (RG) approach predicts
logarithmic violations of the mean field  scaling relations in four
dimensions \cite{BrLeZi76,LW}. It has been stated \cite{ADCCaFr} that
the existence of such logarithmic corrections  to any mean field scaling
relation implies triviality. Hence the importance of the study of these
corrections and the primary motivation for the present work.

The layout of this paper is as follows:

In sect.2 the perturbative RG is applied to the single component
$\phi^4$ theory, and a finite-size scaling (FSS) theory is
developed with particular emphasis on four dimensions. The approach to
criticality from within both the symmetric and broken phases is
studied.

The lattice version of the model is then discussed in sect.3.
The concept of partition function zeroes as an alternative
way to view the onset of a phase transition is recalled.
Formulae describing the FSS behaviour of these zeroes are derived.
These are then used to derive  the FSS formulae of thermodynamic
functions.

The RG and FSS equations predict logarithmic violations of the mean
field theory in four dimensions. These solutions are based
on perturbation theory and have to be tested in an independent
approach.

In sect.4 details of our numerical simulations are discussed
and the results regarding logarithmic scaling corrections presented.
These are consistent with our RG predictions. We conclude in sect.5.

We should mention here that some of the background material is not new.
It has been included to keep the presentation as self contained as
possible. Our treatment of the FSS behaviour of the Lee--Yang zeroes (as
well as the consequent FSS behaviour of the thermodynamic functions)
in four dimensions is, however, new.
Some of our numerical results on Fisher zeroes have been presented
earlier\cite{KeLa91}.

%%%%%%%%%%%%%%%%%%%%%%%%%%%%%%%%%%%%%%%%%%%%%%%%%%%%%%%%%%%%%%%%%%%%%%%%
\section{Renormalization group and finite-size scaling}
%%%%%%%%%%%%%%%%%%%%%%%%%%%%%%%%%%%%%%%%%%%%%%%%%%%%%%%%%%%%%%%%%%%%%%%%
\setcounter{equation}{0}

In the single component version of $\phi^4$ theory with quadratic
composite fields, the Hamiltonian density in $d$-dimensional Euclidean
space-time continuum may be written as \cite{BrLeZi76}
\begin{equation}
{\cal{H}} = \frac{1}{2}(\nabla \phi)^2 +
\frac{m_0^2}{2}\phi^2(x) + \frac{g_0}{4!}\phi^4(x)
-H(x)\phi(x) - \frac{t(x)}{2}\phi^2(x).
\label{BLZ6.36}
\end{equation}
Here $m_0$ is the bare mass of the bosons described by the theory, and
$g_0$ is the bare interaction coupling. $H(x)$ and $t(x)$ are the
sources for the fields $\phi$ and the composite $\phi^2$ fields
respectively. The generating functionals $Z[H,t]$ and $W[H,t]$
are defined by
\begin{equation}
  Z[H,t] = e^{W[H,t]} = C \int \prod_x d \phi (x) e^{-\int
d^dx{\cal{H}}}, \label{R6.1}
\end{equation}
and the constant $C$ is chosen such that
\begin{equation}
 W[0,0] = 0.
\label{normofZ}
\end{equation}
The function conjugate to $H(x)$ is
\begin{equation}
 M(x,t) = \frac{\delta W [H,t]}{\delta H(x)} = \langle \phi(x)
\rangle_{H,t}.
\label{BLZ2.15prime}
\end{equation}
Then the generating functional $\Gamma [M,t]$ of the one particle
irreducible vertex functions is defined through the Legendre
transformation
\begin{equation}
\Gamma[M,t] + W[H,t] = \int dx H(x)M(x),
\label{Legendre}
\end{equation}
with
\begin{equation}
 H(x,t) = \frac{\delta \Gamma [M,t]}{\delta M(x)}.
\label{BLZ2.19}
\end{equation}
The objects of interest are the Schwinger functions (or correlation
functions). Define
\begin{eqnarray}
\lefteqn{ \Gamma^{(L,N)}[y_1,\dots ,y_L;x_1,\dots,x_N;M;t] =}
\nonumber \\
& &
\frac{\delta^{L+N}\Gamma[M,t]}{\delta t(y_1) \dots \delta t(y_L)
      \delta M(x_1) \dots \delta M(x_N)} \quad.
\label{B20}
\end{eqnarray}
This is a functional of $M$ and $t$ (it is a function of its
remaining arguments).

At some value of the bare mass, the renormalized theory is massless.
Setting, then, $m_0$ to this critical value, and letting $t(x)$ become
independent of $x$, $t$ becomes a measure of the deviation away from
the massless theory. If the source $H$ is also independent of $x$, then
the $\Gamma^{(L,N)}$ become functions (rather than functionals) of all
their arguments. This is the situation henceforth assumed.

Power counting yields the (primitive) degree of divergence of an
individual graph. In less than four dimensions
\[ \Gamma^{(0,2)}, \Gamma^{(0,4)} {\rm{ ~ and ~ }} \Gamma^{(1,2)} \]
are divergent.
Renormalization of these three functions gives the renormalized mass
$m_R$, the quartic coupling $g_R$, the field strength renormalization
and the composite field strength renormalization ($Z^{\frac{1}{2}}$ and
$Z_{(2)}$ respectively). The composite field strength is renormalized
by defining $(\phi^2)_R$ as $Z_{(2)} \times  (\phi_R)^2$. Thus
\begin{equation}
(\phi^2)_R = \frac{Z_{(2)}}{Z} \phi^2,
\end{equation}
and the divergence of $Z_{(2)}$ renders $(\phi^2)_R$
finite\cite{BrLeZi76}.

In four dimensions there appears an additional divergence due to
\[  \Gamma^{(2,0)}.  \]
The corresponding diagram has no external legs and can never appear as
a subdiagram. The renormalization of $\Gamma^{(2,0)}$ is therefore
accomplished by subtracting its divergent part. This subtraction
does not effect the other $\Gamma^{(L,N)}$. The relationship between
the bare and the massless renormalized theory is \cite{BrLeZi76,LW}
\begin{eqnarray}
\lefteqn{
 \Gamma_R^{(L,N)}(q_1,\dots ,q_L;p_1,\dots ,p_N;g_R,\mu)
  = \left(\frac{Z_{(2)}}{Z}\right)^L Z^{N/2}}
 \nonumber \\
& &
\times \left\{
\Gamma^{(L,N)}_{\rm{bare}}(q_1,\dots ,q_L;p_1,\dots ,p_N;g_0,\Lambda)
    \right.
           \nonumber \\
& & \quad \quad \quad  \left.
  - \delta_{L,2} \delta_{N,0} \left.
\Gamma_{\rm{bare}}^{(2,0)}(q,-q;g_0,\Lambda)
\right|_{q^2=\frac{4}{3}\mu^2}     \right\}  ,
\label{LWI4.18}
\end{eqnarray}
in which $\Lambda$ is the ultra-violet cutoff and $\mu$ is an arbitrary
non-vanishing mass parameter.

Following the renormalization prescription of \cite{BrLeZi76}, one
can then proceed to expand around this critical (massless) theory.
This allows one
to examine the approach to criticality from within both the symmetric
and the broken phases.

The renormalization group equations express the invariance of the
physics under a rescaling of the mass parameter $\mu \rightarrow \mu
(\lambda)$ defined by
\begin{equation}
\mu(\lambda) \equiv \lambda \mu .
\label{BLZ6.41a}
\end{equation}
Following \cite{BrLeZi76} one finds the RGE
\begin{eqnarray}
\lefteqn{\left[ \lambda \frac{\partial}{\partial \lambda} - \frac{N}{2} \eta
(g_R(\lambda))
 -L \left(  \frac{1}{\nu (g_R(\lambda))} - 2   \right)   \right]
}  \nonumber \\
& &    \times  \Gamma_R^{(L,N)}(q;p;t(\lambda),
M(\lambda),g_R(\lambda),\mu(\lambda)) = \Theta_{L,N}(\lambda)\quad ,
\label{cfBLZ6.42}
\end{eqnarray}
where
\begin{equation}
\Theta_{L,N}(\lambda)  =
\frac{\delta_{N0}}{(2-L)!}
t(\lambda)^{(2-L)} \Upsilon (g_R(\lambda))
\end{equation}
for $L\leq 2$ and zero otherwise. We will refer to this term as the
inhomogeneous term. It comes from the need to (additively) renormalize
$\Gamma^{(2,0)}$ in $d=4$. Since $\Gamma^{(2,0)}$  is not divergent
for $d<4$, the inhomogeneous term in (\ref{cfBLZ6.42}) does not
give rise to any singular contribution there. The flow equations are
\begin{equation}
\begin{array}{rclcrcl}
\di \lambda \frac{d}{d\lambda}g_R(\lambda) &=\D B(g_R(\lambda))
&\mbox{~with~} &       g_R(1) &=& g_R ,\\
\di \frac{\lambda}{t(\lambda)}
 \frac{dt(\lambda)}{d\lambda} &=\D 2 -  \frac{1}{\nu(g_R(\lambda))}
 &\mbox{~with~}  &       t(1) &=& t ,    \\
\di   \frac{\lambda}{M(\lambda)}
 \frac{dM(\lambda)}{d\lambda} &=\D -\frac{1}{2} \eta (g_R(\lambda))
 &\mbox{~with~}  &          M(1) &=& M
\label{BLZ6.41}
\end{array}
\end{equation}
and
\begin{equation}
 \Upsilon (g_R) = \left[2 \left( \frac{1}{\nu(g_R)} - 2 \right) -\mu
\frac{d}{d \mu} \right]
                  \left( \frac{Z_{(2)}}{Z}  \right)^2
                  \left. \Gamma_{\rm{bare}}^{(2,0)} \left(
                  q,-q;g_0, \Lambda
                  \right)\right|_{q^2 = \frac{4}{3}\mu^2}.
\label{BLZ6.32}
\end{equation}
%%%%%%%%%%%%%%%%%%%%%%%%%%%%%%%%%%%%%%%%%%%%%%%%%%%%%%%%%%%%%%%%%%%%%%%%
The content of the RGE (\ref{cfBLZ6.42}) is, then, that if $\mu$ is
rescaled by a factor $\lambda$ then the response of $g_R$, $t$ and $M$
is governed  by (\ref{BLZ6.41}).

The Callan Symanzik beta function \cite{CaSy70} is denoted by $B$ in
(\ref{BLZ6.41}). The form of this function can be calculated
perturbatively in the renormalized $d=(4-\epsilon )$ dimensional theory.
Eq.(\ref{BLZ6.41}) then gives the behaviour of the running coupling
constant $g_R(\lambda)$. It turns out that in order to remove the
cutoff, the running coupling constant $g_R(\lambda)$ has to approach the
infra--red (IR) fixed point $g_R^* \sim O(\epsilon)$.
This fixed point then governs the critical region.

The region of interest can be divided into  \\
(a) $t \ge 0$, $H=0$. In this region there is no magnetization
( $M=0$ ). This is a symmetric theory. \\
(b) $t>0$ and $H\neq0$ or $t<0$ and any $H$.
This is the region of broken symmetry ( in which $M \neq 0$).\\
The RGE (\ref{cfBLZ6.42}) holds throughout regions (a) and (b).
The point in using vertex functions is that it is nowhere necessary
to state whether or not the symmetry is broken (be it spontaneously
or explicitly) \cite{Amit}.

The solution of (\ref{cfBLZ6.42}) is
\begin{eqnarray}
\lefteqn{\Gamma_R^{(L,N)}(q;p;t,M,g_R,\mu)
=  \tilde{Z}(\lambda,g_R(\lambda))  } \nonumber \\
& &  \times \Gamma_R^{(L,N)}
\left(q;p;t(\lambda),M(\lambda),g_R(\lambda),\mu(\lambda) \right)
  + \Pi_{L,N}(\lambda),
\label{soln}
\end{eqnarray}
where
\begin{eqnarray}
    \tilde{Z}(\lambda,g_R(\lambda)) & = &
    e^{- \int_{g_R}^{g_R(\lambda)} \left[ \frac{N}{2} \eta (g_R) +
    L \left(  \frac{1}{\nu (g_R)} - 2 \right)
\right]\frac{dg_R}{B(g_R)}} \nonumber \\
& = &
 \left( \frac{M(\lambda)}{M}  \right)^N
 \left( \frac{t(\lambda)}{t}  \right)^L ,
\label{cfBLZ6.47b}
\end{eqnarray}
and the inhomogeneous term is
\begin{equation}
 \Pi_{L,N}(\lambda) = \int_1^\lambda
   \frac{d\lambda^\prime}{\lambda^\prime}
   \Theta_{L,N}(\lambda^\prime)
   \tilde{Z}(\lambda^\prime,g_R(\lambda^\prime)).
\end{equation}

Because of the local nature of the RG, the renormalization constants
of the infinite volume theory render finite the finite volume
theory too \cite{Br82,Ca88b}. We denote by
$\Gamma_R^{(L,N)}(q;p;t,M,g_R,\mu,l)$ the renormalized Schwinger
function of the finite volume theory,  where $l$ denotes the linear
extent of the system.
The  RGE obeyed by this Schwinger function is the same as
(\ref{cfBLZ6.42}). Its solution is (from (\ref{soln}))
\begin{eqnarray}
\lefteqn{
  \Gamma_R^{(L,N)}(q;p;t,M,g_R,\mu,l)
=  \tilde{Z}(\lambda,g_R(\lambda))
}
      \nonumber \\
& & \times \Gamma_R^{(L,N)}
\left(q;p;t(\lambda),M(\lambda),g_R(\lambda),\mu \lambda,l \right)
  + \Pi_{L,N}(\lambda).
\label{soln2}
\end{eqnarray}
To prepare for dimensional analysis, we implicitly replace $g_R$ by
$\mu^{\epsilon} g_R$ to keep it dimensionless. Then applying dimensional
analysis to the homogeneous term on the right hand side gives
\begin{eqnarray}
\lefteqn{
          \Gamma_R^{(L,N)}(q;p;t,M,g_R,\mu,l)
          =  \tilde{Z}(\lambda,g_R(\lambda))
          l^{ \frac{N}{2} (d-2) +2L-d }
}
      \nonumber \\
& &  \times \Gamma_R^{(L,N)}
  \left(
         lq;lp;l^2t(\lambda),l^{\frac{d-2}{2}}M(\lambda),g_R(\lambda),
  l\mu \lambda,1
  \right)
  + \Pi_{L,N}(\lambda).
\label{Br15generalized}
\end{eqnarray}
Since $\lambda$ is still at our disposal (as long as it is small enough
so as to remain in the critical region),
we choose
\begin{equation}
 l \mu \lambda  =1.
\end{equation}
Then,
\begin{eqnarray}
\lefteqn{  \Gamma_R^{(L,N)}(q;p;t,M,g_R,\mu,l) }
\nonumber \\
& & =  \left( \frac{M(1/l\mu)}{M}
    \right)^N
   \left( \frac{t(1/l\mu)}{t} \right)^L
    l^{\frac{N}{2}(d-2)+2L-d}
\nonumber \\
& & \times \Gamma_R^{(L,N)}\left(lq;lp;
  l^2t\left(\frac{1}{l\mu}\right),
   l^{\frac{d-2}{2}}M\left(\frac{1}{l\mu}\right),
   g_R\left(\frac{1}{l\mu}\right),
   1,1\right)
\nonumber \\
& &  + \Pi_{L,N}\left(\frac{1}{l\mu}\right).
\label{my2.17}
\end{eqnarray}
In less than four dimensions, and in the critical region,
the flow equations give \cite{BrLeZi76,Br82}
\begin{equation}
t\left(\lambda\right) = t \lambda^{(2- \frac{1}{\nu})}
\end{equation}
\begin{equation}
 M\left(\lambda\right) = M \lambda^{-\frac{1}{2}\eta}.
\end{equation}
If $l\mu$ is large enough then
$g_R(\frac{1}{l\mu})$ is close to $g_R^*$.
Therefore, at zero momentum,
\begin{eqnarray}
\lefteqn{ \Gamma_R^{(L,N)}(0;0;t,M,g_R,\mu,l)
=    \mu^{\frac{N}{2}\eta + L\left(\frac{1}{\nu}-2 \right)}
      l^{\frac{N\beta}{\nu}+\frac{L}{\nu}-d}}
\nonumber \\
& &
  \times \Gamma_R^{(L,N)}
    \left(0;0;\mu^{\frac{1}{\nu}-2} l^{\frac{1}{\nu}} t,
              \mu^{\frac{1}{2}\eta}l^{\frac{\beta}{\nu}}M,
              g_R^*,1,1  \right)
  + \Pi_{L,N}\left(\frac{1}{l\mu}\right),
\end{eqnarray}
where
\begin{equation}
\beta = \frac{\nu}{2}(d-2+\eta).
\end{equation}
If $\mu$ is fixed, then
\begin{equation}
  \Gamma_R^{(L,N)}(0;0;t,M,g_R,\mu,l)
  =      l^{\frac{N\beta}{\nu}+\frac{L}{\nu}-d}
    F_{\mu}^{(L,N)}
    \left(l^{\frac{1}{\nu}}t,l^{\frac{\beta}{\nu}}M \right)
  + \Pi_{L,N}\left(\frac{1}{l\mu}\right),
\end{equation}
where $F_\mu^{(L,N)}$ is an unknown function of its arguments.
Eq.(\ref{BLZ2.19}) can be applied to this form for
$\Gamma_R^{(0,0)}$ to express the external field $H$ in terms of $M$.
This gives
\begin{equation}
  \Gamma_R^{(L,N)}(0;0;t,H,g_R,\mu,l)  =
  l^{\frac{N\beta}{\nu} + \frac{L}{\nu}-d }
  {F^\prime_\mu }^{(L,N)} \left(l^{\frac{1}{\nu}}t,l^{\frac{\delta
\beta}{\nu}} H  \right)
  + \Pi_{L,N}\left(\frac{1}{l\mu}\right).
\label{my2.19}
\end{equation}
Here $\delta$ is  the usual odd critical exponent
defined by
\begin{equation}
 \delta = \frac{d+2-\eta}{d-2+\eta}.
\end{equation}
Eq.(\ref{my2.19}) is sufficient to derive the usual FSS relations in
less than four dimensions \cite{Fi72,Br82,Ba83}.

For example, the zero field susceptibility is given by
\[
  \chi_l^{-1}(t)  =  \Gamma_l^{(0,2)}
    \left(0;t,0,g_R,\mu \right)
   =  l^{\frac{2\beta}{\nu}-d}
 F_\chi (l^{\frac{1}{\nu}} t),
\]
where $F_\chi$ is, again,  an unknown function.
Putting $t=0$ then gives the FSS
behaviour of the zero field susceptibility at the infinite volume
critical point, $\chi_l(t=0) \propto l^{2-\eta}$.

In the four dimensional version of the theory there appear certain
subtleties which  are not present below four dimensions.
This is because the IR fixed point of the
Callan-Symanzik function $B(g_R)$ moves to the origin as the dimension
becomes four. Secondly, in contrast to the $d<4$ dimensional case,
the fixed point is now a double zero, responsible for the occurrence of
logarithmic corrections.

A third difference between the cases of $d<4$ and $d=4$ comes from
the inhomogeneous term in the RGE. The graph responsible for this
term is not in fact divergent when $d<4$. Singular behaviour
in less than four dimensions comes from the homogeneous term. In $d=4$
the inhomogeneous term contributes to the leading singular behaviour
too. The first term remains singular however, and is responsible for
divergences such as that in the susceptibility.

Eq.(\ref{my2.19}), from which the FSS behaviour of the model can be
derived below four dimensions, was established with the help of the
approximation $g_R(\frac{1}{\mu l}) \simeq g_R^*$ for large $\mu l$.
As pointed out by Br\'ezin in \cite{Br82}, this approximation fails in
four  dimensions. The reason is that $g_R^*$ then becomes zero, and one
is left with the mean field theory.

In $d=4$, we then have to rely on a perturbative expansion in $g_R$.
To lowest order, the functions
$B(g_R)$, $\eta(g_R)$, $\nu(g_R)$ and $\Upsilon (g_R)$
are\cite{BrLeZi76,LW}
\begin{eqnarray}
  B(g_R) & = & \frac{3}{2} g_R^2 -\frac{17}{12}g_R^3     + O(g_R^4),
\label{BLZ8.13aaa} \\
\eta (g_R) & = & \frac{1}{24} g_R^2 + O(g_R^3),
\label{BLZ8.13b} \\
  \frac{1}{\nu(g_R)} & = & 2 - \frac{1}{2} g_R + O(g_R^3),
\label{BLZ8.13c} \\
  \Upsilon (g_R) & = & \frac{1}{2} + O(g_R).
\label{BLZ8.22}
\end{eqnarray}
Putting $\mu=1$ for simplicity, these perturbative solutions, together
with the flow equations (\ref{BLZ6.41}), give for (\ref{my2.17})
\begin{eqnarray}
  \Gamma_R^{(L,N)}\left(q;p;t,M,g_R,1,l\right)
   \simeq
    \left(
     \frac{2}{3g_R\ln{ l}} \right)^{L/3}
    l^{N+2L-4}
    \nonumber  \\
  \times
  \Gamma_R^{(L,N)}\left(lq;lp;l^2t\left( \frac{2}{3g_R\ln{l}}  \right)^{1/3},
                   lM,\frac{2}{3\ln{l}},1,1\right)
     \nonumber \\
  + \frac{\delta_{N0}}{(2-L)!}\frac{3}{2}
    \left( \frac{2}{3g_R}  \right)^{2/3}
     t^{2-L}
   \left( \ln{ l}  \right)^{1/3}. ~ ~ ~ ~ ~ ~ ~
\label{Hl202}
\end{eqnarray}
The coupling constant for the Schwinger function on the right hand side
is $\frac{2}{3\ln{l}}$ for large $l$. Since this is small, perturbation
theory may be applied to calculate $\Gamma_R^{(0,0)}$\cite{BrLeZi76}.
This gives
\begin{eqnarray}
\lefteqn{\Gamma_R^{(0,0)}\left( t,M,g_R,1,l  \right) }
\nonumber \\
& & = c_1 tM^2 \left( \ln{l}  \right)^{-1/3}   +
   c_2 M^4 (\ln{l})^{-1}   +
   c_3 t^2 (\ln{l})^{1/3}
\label{Gamma00bkn}
\end{eqnarray}
where $c_1,c_2$ and $c_3$ are constants.
Applying (\ref{BLZ2.19}) to this yields for the external field
\begin{equation}
H\left( t,M,g_R,1,l  \right)  \simeq
 c_4 t M (\ln{l})^{-1/3}  +  c_5 M^3 (\ln{l})^{-1},
\label{Hl300}
\end{equation}
where, again, $c_4$ and $c_5$ are constants.

The free energy per unit volume in the presence of an external field
is
\begin{equation}
  W_l(t,H) = M H(t,M;l) - \Gamma_R^{(0,0)}(t,M;l).
\label{W1stway}
\end{equation}
Eqs.(\ref{Hl300}) and (\ref{Gamma00bkn}) give, then,
\begin{equation}
 W_l(t,H) = c_1^\prime \frac{tM^2}{(\ln{l})^{1/3}}  +
   c_2^\prime \frac{M^4}{\ln{l}} +
   c_3  t^2 (\ln{l})^{1/3},
\label{WtHl4D}
\end{equation}
where $c_1^\prime$ and $c_2^\prime$
are constants and $M$ is related to $H$ through
(\ref{Hl300}).
This expression is the basis of all the FSS relations derived below.

If $H$ vanishes, then all  of the solutions of (\ref{Hl300}) lead to
\begin{equation}
  W_l(t,0) \propto t^2 \left( \ln{l}  \right)^{\frac{1}{3}}.
\label{Wt0l4D}
\end{equation}

%%%%%%%%%%%%%%%%%%%%%%%%%%%%%%%%%%%%%%%%%%%%%%%%%%%%%%%%%%%%%%%%%%%%%%%%
\section{Lattice $\phi^4$ theory and the zeroes of the partition
function }
%%%%%%%%%%%%%%%%%%%%%%%%%%%%%%%%%%%%%%%%%%%%%%%%%%%%%%%%%%%%%%%%%%%%%%%%
\setcounter{equation}{0}

Within the path integral formulation of quantum field theory there are
two complimentary approaches. The first is perturbation theory (in the
quartic coupling $g_R$). Indeed this is the basis for the considerations
at the end of the previous section. The second approach is intrinsically
non-perturbative. It involves the use of stochastic techniques to
calculate the path integrals. Apart from statistical errors
numerical approaches are exact, but limited to finite lattice volumes.

We have used such a numerical approach --- the Monte Carlo (MC) method
--- to further study the logarithmic corrections involved in four
dimensions. In particular, we present numerical evidence of the
validity of the FSS
formulae  presented in the last section. Thus we have two independent
approaches, whose agreement leaves little doubt that this FSS analysis
indeed correct. This provides support for the validity of the
analyses presented in \cite{BrLeZi76},\cite{LW} and \cite{Br82}, and for
the triviality of $\phi_4^4$ theory.

The usual regularization for a numerical approach replaces
the space-time continuum by a lattice. This is, of course, entirely
equivalent to the use of the momentum cut-off in sect.2.
We use a regular hypercubic lattice of unit intersite spacing. If $t$ is
independent of $x$ in (\ref{BLZ6.36}), the lattice parameterized action
in the absence of a source field and with finite differences replacing
derivatives reads
\begin{equation}
  -\kappa \sum_{x,\mu}\phi_x \phi_{x+\mu}   +   \sum_x \phi_x^2
 + \lambda \sum_x\left( \phi_x^2-1  \right)^2 \quad .
\label{Lang2.6}
\end{equation}
Here the hopping parameter $\kappa$ and the quartic coefficient
$\lambda$ correspond, in a sense, to the mass and quartic coupling of
the continuum theory respectively. Taking $\lambda$ to infinity gives
the Ising limit of the model. Here, the fields $\phi_x$ take only values
from the set $\{\pm 1 \}$. The universality hypothesis, which comes from
experience in statistical physics, implies that no information should
be lost in going to the Ising extreme. I.e., the Ising model and the
$\phi^4$ model with arbitrary $\kappa$ and $\lambda$ should be in the
same universality class and exhibit the same scaling behaviour (for a
related MCRG study cf.\cite{La86}). The vacuum to vacuum transition
amplitude of the quantum field theory becomes the partition function of
the Ising model.

In the presence of an external field the Ising model can be defined by
the partition function
\begin{equation}
Z(\kappa,H) = \frac{1}{\cal{N}} \sum_{\{\phi\}} e^{\kappa S + hM}
\end{equation}
where
\begin{equation}
 S = \sum_x \sum_{\mu=1}^{d} \phi_x \phi_{x+\mu}
\quad , \quad  M = \sum_x \phi_x \quad.
\end{equation}
Here, the Boltzmann factor has been absorbed into the hopping
parameter $\kappa$, and into the reduced external field
$h = \kappa \times H$. The sum runs over all $\cal{N}$ possible
configurations of the spin field on the $d$ dimensional lattice, and the
normalization ensures $Z(0,0)=1$. We may reexpress the partition
function by
\begin{eqnarray}
Z(\kappa,h) &=& \sum_{M=-N}^{N} \sum_{S=-dN}^{dN}
  \rho(S,M) e^{\kappa S + hM} \nonumber \\
 &=& \sum_{M=-N}^{N} \rho(\kappa;M) e^{hM}
 = \sum_{S=-dN}^{dN}   \rho(S;h) e^{\kappa S} \quad,
\label{Znum}
\end{eqnarray}
where $N$ is the number of sites on the lattice.
The spectral density $\rho(S,M)$ denotes the relative weight of
configurations having given values of $S$ and $M$. By $\rho(\kappa;M)$
and $\rho(S;h)$ we denote the correspondingly integrated densities.

$Z$ is a polynomial in the fugacity $e^{2h}$ (degree $N$) and
$e^{4\kappa}$ (degree $dN/2$). The coefficients of the polynomial in
$e^{2h}$ for real constant $\kappa$ are real and positive, as are those
of the polynomial in $e^{4\kappa}$ for real constant $h$. A knowledge of
the zeroes of the partition function is equivalent to a knowledge of $Z$
itself (and of all functions derivable from it).
In particular, the critical behaviour of Ising-type systems can be
analysed through its partition function zeroes instead of more
traditional methods involving real parameters.

The study of partition function zeroes in general
was initiated by Yang and Lee in 1952 \cite{LeYa52}.
The Lee--Yang theorem states that for ferromagnetic systems
all of the zeroes of the partition function in the external ordering
magnetic field variable
lie on the imaginary axis for real temperatures. Fisher was the first to
analyse the zeroes in the complex temperature (or mass) plane
\cite{Fi64}.  Thus we refer to partition function zeroes in the
temperature plane as Fisher zeroes and to those in the complex plane of
external fields as Lee--Yang zeroes. With the exception of systems which
are  self dual \cite{Ma83b}, there exist no simple general results of
the Lee--Yang type concerning the locus of Fisher zeroes.
Thus the vast majority of studies have been of a numerical nature
(see, however \cite{Abe,Suzuki} and references therein).

Itzykson, Pearson and Zuber \cite{ItPeZu} initiated the study of FSS of
partition function zeroes. Their analysis was confined to less than four
dimensions with power-law scaling behaviour and corrections.
This was later extended to dimensions above (not including) four in
\cite{GlPrSc87}. The latter is also restricted to purely power-law
scaling behaviour. In this section, the corresponding FSS theory is
presented for four dimensions where logarithmic corrections are
manifest.

Denote by $C_l(t)$ and $\chi_l(t)$ the specific heat and magnetic
susceptibility (per unit volume) respectively of a system of linear
extent $l$ and at a reduced temperature $t$ in zero external magnetic
field (cf. sect.3). Twice differentiating the free energy in the
perturbative RG formula (\ref{Wt0l4D}) gives
\begin{equation}
 C_l(t) \propto \left( \ln{l} \right)^{\frac{1}{3}}.
\label{fssCl}
\end{equation}

The total free energy  at the critical temperature in four dimensions in
the presence of an  external field is given by (\ref{WtHl4D}) as
\begin{equation}
   l^4  (\ln{l})^{\frac{1}{3}} H^{\frac{4}{3}}.
\end{equation}
The partition function is therefore
\begin{equation}
 Z_l(t=0,H) = Q\left(    l^4  (\ln{l})^{\frac{1}{3}} H^{\frac{4}{3}}
 \right).
\end{equation}
If at some (complex) value of $H$ the partition function vanishes,
then, for this value of $H$,
\begin{equation}
 H^{\frac{4}{3}} = l^{-4} (\ln{l})^{-\frac{1}{3}}
 Q^{-1}(0).
\end{equation}
Therefore
\begin{equation}
 H_j \propto l^{-3} (\ln{l})^{-\frac{1}{4}}
\label{LYzeroesd=4}
\end{equation}
where the constant of proportionality depends on the index $j$ of the
zero. This is the FSS formula for Lee--Yang zeroes in four dimensions.

Eq.(\ref{Wt0l4D}) gives for the total free energy (when $H=0$)
\begin{equation}
  F_l(t,H=0) \propto
  l^4 t^2 \left( \ln{l}  \right)^{\frac{1}{3}}.
\end{equation}
The partition function is the exponential of this, i.e.,
\begin{equation}
   Z_l(t,H=0) =
    R\left(l^4  t^2 \left( \ln{t} \right)^{1/3}\right).
\end{equation}
If $R$ vanishes, then,
\begin{equation}
 t^2 l^4 \left( \ln{t} \right)^{1/3} = R_j^{-1}(0)
\end{equation}
where $j$ indicates the index of the zero.
Therefore, for
the  $j^{{\rm{th}}}$ zero,
\begin{equation}
 t_j \propto  l^{-2} \left( \ln{l} \right)^{-1/6}
\label{fssfisher4D}
\end{equation}
where the proportionality constant depends, again, on $j$.

The scaling relations for the partition function zeroes
can be used to find the   behaviour of the thermodynamic functions
as well. The partition function is a polynomial and as such can be
written in terms of its zeroes. Let $H_j$ be the $j^{\rm{th}}$ Lee--Yang
zero for a system of linear extent $l$. Then, the partition function is
\begin{equation}
 Z_l(\kappa,H) \propto \prod_j{\left( H-H_j \right)}.
\end{equation}
The magnetic susceptibility is given by the second derivative of the
Gibbs free energy
with respect to $H$. This gives, in $d=4$,
\begin{equation}
 \chi_l(\kappa,H) \propto \frac{1}{l^4}
 \sum_j{\frac{1}{\left(H-H_j\right)^2}}.
\end{equation}
Therefore the susceptibility  at the critical value of $H$ (namely
at $H=0$) is
\begin{equation}
 \chi_l \left( \kappa,0 \right)
 \propto \frac{1}{l^4} \sum_j{ \frac{1}{H_j^2} } .
\end{equation}
Eq.(\ref{LYzeroesd=4}) then gives
the FSS formula for the zero field susceptibility in four dimensions
as
\begin{equation}
 \chi_l\left(\kappa_c,H=0\right)
 \propto
 l^2 \left( \ln{l} \right)^{\frac{1}{2}}.
\label{chil4D}
\end{equation}

A similar calculation for the Fisher zeroes leads to the recovery of the
FSS formula for specific heat (\ref{fssCl}). Let $\kappa_j$ be the
$j^{\rm{th}}$ Fisher zero for a system of linear extent $l$.
Then, in zero field, the partition function is
\begin{equation}
 Z_l(\kappa) \propto \prod_j{\left( \kappa-\kappa_j \right)}.
\end{equation}
The specific heat is given by the second derivative of the free energy
with respect to $t$. This gives, in $d=4$,
\begin{equation}
 C_l(\kappa) = -\frac{1}{l^4}
 \sum_j{\frac{1}{\left(\kappa-\kappa_j\right)^2}}.
\end{equation}
Therefore the specific heat at the critical value of $\kappa$ is
\begin{equation}
 C_l \left( \kappa_c \right)
 = -\frac{1}{l^4} \sum_j{ \frac{1}{\tau_j^2} }
\end{equation}
where $\tau_j$  is the `reduced' position of the
$j^{\rm{th}}$ zero:
\begin{equation}
 \tau_j = \kappa_j - \kappa_c.
\end{equation}
Eq.(\ref{fssfisher4D}) gives
\begin{equation}
 \tau_j \propto l^{-2}\left( \ln{l} \right)^{- \frac{1}{6}}
\end{equation}
where the constant of proportionality depends on the index $j$.
Thus the FSS formula for the specific heat in four dimensions
is \cite{RuGuJa85}
\begin{equation}
 C_l\left(\kappa_c\right)
 \propto
 \left( \ln{l} \right)^{\frac{1}{3}}.
\label{Cl4D}
\end{equation}

The partition function zeroes provide an alternative way to view the
onset of criticality. As the system size increases towards infinity, the
zeroes tend to pinch the real $H$ or $\kappa$ axes ( (\ref{LYzeroesd=4})
and (\ref{fssfisher4D})). Thermodynamic observables such as the specific
heat and magnetic susceptibility become divergent. This applies to the
correlation length as well. The FSS formula for the correlation
length of a four dimensional system also involves logarithmic
corrections. This was derived by Br\'ezin \cite{Br82} for a system of
extent $l$ in all  directions. At the infinite volume critical point
$\kappa = \kappa_c$, one has
\begin{equation}
      \xi_l(\kappa_c ) \propto l (\ln{l})^{\frac{1}{4}}
\end{equation}
This suggests that a FSS variable should indeed be defined by
\begin{equation}
\frac{\xi_\infty(\kappa)}{\xi_l(\kappa_c)} =
\frac{t^{-\ot} \mid\;\ln t\mid\;^{\frac{1}{6}} }{l
(\ln{l})^{\frac{1}{4}}} \end{equation}
in four dimensions\cite{KeLa91}.

%%%%%%%%%%%%%%%%%%%%%%%%%%%%%%%%%%%%%%%%%%%%%%%%%%%%%%%%%%%%%%%%%%%%%%%%
\section{Numerical calculations and results}
%%%%%%%%%%%%%%%%%%%%%%%%%%%%%%%%%%%%%%%%%%%%%%%%%%%%%%%%%%%%%%%%%%%%%%%%
\setcounter{equation}{0}

We now want to report on our numerical calculations which confirm the
scaling picture of sect.2. In particular, we want to identify the
multiplicative logarithmic corrections  to FSS. Such logarithmic
corrections have been notoriously difficult to verify numerically (see
e.g. \cite{ADCCaFr} and \cite{RuGuJa85}). However, the advent of more
efficient cluster algorithms \cite{SwWa87,Wo89} has greatly improved the
quality of Monte Carlo calculations for bosonic spin systems like the
Ising model. We suggest --- and the quality of our results supports our
proposal --- that a study of the FSS of partition function zeroes lends
itself more readily to the detection of logarithms than do the more
traditional thermodynamic quantities such as specific heat.

The first numerical calculations of partition function zeroes appeared
in the 1960's \cite{OnKaSuKa67}. Such early work involved exact
calculations of the density of states (spectral density) $\rho(S,M)$ and
were therefore confined to very small lattices. (See \cite{KaAbYa71} for
a list of references and early history).

The next major step concerning numerical calculations was made by
Falcioni et al.\cite{FaMaPa82} and by Marinari et al.\cite{Ma84} in the
early 1980's. They were the first to use approximations to the density
of states in the form of histograms to study critical phenomena. It is
clear that straightforward analytical continuation can take (\ref{Znum})
to the complex $\kappa$ or $H$ plane. Thus the histogram technique can
be used for a precise numerical determination of the complex partition
function zeroes.

Numerical methods received a further boost with the development of
techniques whereby a number of Monte Carlo constructed histograms can be
combined to form one `multihistogram' \cite{FeSw88,AlBeVi90a}.
These provide a better approximation to the spectral density over a
wider range  of the parameter $\kappa$ or $H$.

We now present some details of our own numerical calculations. The data
were taken on lattices of size from $8^4$ to $24^4$ using the
Swendsen--Wang cluster algorithm. Histograms were determines at $h=0$
and at various values of $\kappa$ close to the pseudocritical one
(chosen to be that value of $\kappa$ where the specific heat peaks).
Table 1 provides a list of lattices sizes, the values of $\kappa$ at
which the simulations took place, as well as a summary of the
statistics.

For the determination of $\rho(S;h=0)$ the various `raw' histograms were
suitably combined following \cite{FeSw88}; no binning was used. With
(\ref{Znum}) this allows one to construct $Z(\kappa,h=0)$ in the complex
neighbourhood of the real $\kappa$-values and to determine nearby Fisher
zeroes.

For $\rho(\kappa; M)$ we binned each of the raw $(S,M)$-histograms in a
$256\times256$ array and then combined for each $M$-bin the
corresponding $S$-subhistograms according \cite{FeSw88}. This then
allows us to obtain an optimal $\rho(\kappa_0;M)$ for arbitrary
$\kappa_0$ in the considered domain. From this $Z(\kappa_0, h)$ may be
determined for not too large values of (imaginary) $h$. Below we present
results for Lee--Yang zeroes evaluated at $\kappa_0 = \kappa_c$. As a
consistency check we also determined the zeroes coming from
a single $M-$histogram corresponding to a simulation at $\kappa_c$.

The errors in the quantities calculated from
the multihistograms were estimated by the jackknife method, i.e. the
data for each lattice size were cut to produce $10$
subsamples leading to different multihistograms and thus to different
results, whence the variance and bias were calculated \cite{Bo89}.

{}From the Lee--Yang theorem \cite{LeYa52} it is known that the zeroes in
$H$ all lie on the imaginary axis for any lattice size $l$. The search
for the Lee--Yang  zeroes is therefore technically easier than for the
Fisher zeroes. In the later case we used a Newton--Raphson type
algorithm. In order to avoid instabilities due to the large numbers
involved, and since $Z_l(\kappa)$ never vanishes for real $\kappa$, the
Fisher zeroes were in fact found as local steep minima in
$\left|Z_l(\kappa ) / Z_l({\rm{Re ~ }} \kappa) \right|^2$.

We now come to the FSS analysis of the Fisher zeroes. The positions of
the closest two Fisher zeroes obtained from the multihistograms are
listed  in table 2 (where $\kappa_j$ represents the $j^{\rm{th}}$ Fisher
zero). Since we can confine the scaling analysis to the imaginary parts
of the zeroes we avoid the necessity of knowing the infinite volume
critical value of $\kappa$. In fig.1a we plot the logarithm of the
imaginary part of the position of the first Fisher zero
against the logarithm of the lattice size $l$. A linear fit to the slope
($-\frac{1}{\nu }$) gives $\nu = 0.479(1)$ which is slightly below the
mean field value of $\ot$. This deviation from the mean field value is
due to the presence of logarithmic corrections, which we have neglected
in this first fit. A corresponding analysis applied to the second Fisher
zeroes gives $\nu = 0.467(8)$.

Assuming that the leading scaling behaviour is indeed proportional
to $ l^2$, we can proceed to search for
multiplicative logarithmic corrections. To this end, we plot in fig.1b
$\ln{(l^2 {\rm{Im}} \kappa_1)}$ versus $\ln{(\ln{l})}$. A negative slope
is clearly identified and is in good agreement with the scaling
prediction of $- \frac{1}{6}$. In fact, a fit to all five points gives a
slope $-0.217(12)$. Excluding the point corresponding to $l = 8$ gives
a slope of $-0.21(4)$. The solid line is the best fit to the points
corresponding to $l=12\ldots 24$ assuming the theoretical prediction
$-\frac{1}{6}$ from (\ref{fssfisher4D}).

The errors in the second (and higher index) Fisher zeroes are too large
to warrant a corresponding analysis.

Now that the logarithmic corrections to the FSS behaviour of the
Fisher zeroes have been established, we may proceed to determine
the infinite volume critical hopping parameter $\kappa_c$ from
\begin{equation}
  \left| \kappa_j - \kappa_c\right|
  \propto
  l^{-2} \left( \ln{l} \right)^{- \frac{1}{6}}.
\end{equation}
Using the first Fisher zeroes, we find $\kappa_c \simeq 0.149703(15)$
in good agreement with the value $0.149668(30)$ from high temperature
expansions \cite{GaSyMc79}.

Mean field theory \cite{ItPeZu} predicts that the angle $\varphi$
at which the Fisher zeroes depart from the real axis should
be $\frac{\pi}{4}$. There exists, unfortunately, no FSS theory for this
quantity. We list, in table 3, our measurements of this angle
defined by the first and second Fisher zeroes and the real axis for
the five lattice sizes analysed and plot it in fig.2.
The average value compares well with the mean field prediction.

Let us now discuss the results for the Lee--Yang zeroes. For all real
$\kappa$ they have to lie on the imaginary $h$-axis. At $\kappa_c$ they
should scale according to the FSS formula (\ref{LYzeroesd=4}).
Table 4 lists the positions of the first two Lee--Yang zeroes
as obtained from the multihistogram at $\kappa_c = 0.149703$.
Fig.3a is a log-log plot of the imaginary parts of the positions of the
first Lee--Yang zeroes against the lattice size. The resulting slope
is $-3.083(4)$; this compares well with mean field prediction of
$-3$. The deviation, again, may be explained by logarithmic corrections.

To identify the logarithmic corrections, we plot in fig.3b
$\ln{( l^3 {\rm{Im}} h_1)}$ against $\ln{(\ln{l})}$.
For the first Lee--Yang zeroes a best fit to all five points gives a
slope of $-0.204(9)$ which compares well with the theoretical prediction
of $-\frac{1}{4}$ from (\ref{LYzeroesd=4}). Excluding the smallest
lattice, a best fit to the remaining four points gives a slope
$-0.22(3)$. The solid line in  fig.3b is the best fit to the last four
point with given slope $-\frac{1}{4}$.

The errors in the positions of the second nearest Lee--Yang zeroes are
only about twice that of the corresponding first index zeroes and these
can also be used to analyse FSS. The results again are
good agreement with the expected scaling behaviour.

%%%%%%%%%%%%%%%%%%%%%%%%%%%%%%%%%%%%%%%%%%%%%%%%%%%%%%%%%%%%%%%%%%%%%%%%
\section{Conclusions}
%%%%%%%%%%%%%%%%%%%%%%%%%%%%%%%%%%%%%%%%%%%%%%%%%%%%%%%%%%%%%%%%%%%%%%%%
\setcounter{equation}{0}

We have used RG techniques to derive the FSS behaviour of the $\phi^4$
theory in d=4, placing particular emphasis on the partition function
zeroes.  These formulae were then tested in a non-perturbative
fashion --- with high precision numerical methods.
Of primary interest are the multiplicative logarithmic
corrections to the leading power law scaling behaviour.
These logarithmic corrections are clearly identified
from the scaling behaviour of the closest Fisher and Lee--Yang zeroes
and are in good quantitative agreement with the theory.
Higher index partition function zeroes and thermodynamic observables
such as the specific heat exhibit logarithmic corrections too.

The high precision of the numerical results and the good quantitative
agreement with the analytical predictions of sect.3 provide
non-perturbative evidence for the existence of a double zero in the
Callan-Symanzik beta function. The fixed point  responsible for
non-trivial behaviour in $d<4$ dimensions has moved to the the position
of the trivial fixed point in d=4. The results support the assertion
that $\phi^4$ theory is trivial in four dimensions.
\\
{}~ \\
{}~ \\

\noindent
{\bf Acknowledgement:}
We wish to thank B. Berg, P. Butera, Ch. Gattringer, H. Gausterer,
M. Salmhofer and T. Trappenberg  for discussions.
\newpage

%---------------------------------------------------------

\newpage
%------------------------------------------------------------------------------
\section*{Tables}
%------------------------------------------------------------------------------

{\noindent \bf Table 1:}
For each lattice size $L$ we list the values of $\kappa$ at which
MC simulations were performed. The corresponding $h$ value is always
zero. In parentheses we give the number of
measurements in units of 1000 (\#/1000).
\smallskip
\begin{center}
\begin{tabular}{|r|llll|}
\hline
    $L$ & $\kappa_0$(\# /1000)& & &\cr
\hline
 8 & 0.149709 (400), &  0.1506 (200), &  0.1515 (200), &  0.1520 (200),   \cr
   &  0.1525 (400), &  0.1527 (400), &  0.1529 (400), &  0.1531 (400),   \cr
   &  0.1533 (400), &  0.1540 (200) & &      \cr
\hline
12 & 0.149709 (200), &  0.1498 (100), &  0.1503 (100), &  0.1508 (200),  \cr
   & 0.1510 (200), &  0.1512 (200), &   0.1514 (200), &   0.1520 (100)  \cr
\hline
16 & 0.1492 (50), &   0.1496 (50), &  0.149709 (150), &   0.1500 (50),  \cr
   & 0.1503 (100), &  0.1504 (100), &  0.1505 (150), &  0.15054 (150),
   \cr
   & 0.1506 (150), &  0.1509 (100), &   0.1511 (50), &    0.1513 (50),
      \cr
   & 0.1518 (50) & & & \cr
\hline
20 & 0.1495 (30), &  0.149709 (80), &  0.1498 (30), &   0.1499 (50),    \cr
   & 0.1500 (50), &  0.1501 (100), &  0.1502 (80), &  0.1503 (80),       \cr
   & 0.1504 (50), &  0.1507 (30) & &                                     \cr
\hline
24 & 0.1495 (20), &  0.149709 (80), &  0.1498 (108), &  0.1499 (68),    \cr
   & 0.1500 (80), &  0.1501 (80), &  0.1502 (80), &  0.1504 (20)     \cr
\hline
\end{tabular}
\end{center}

\vspace{0.5cm}
{\noindent \bf Table 2:}
The positions of the first and second Fisher zeroes as obtained from the
jackknifed multi-histograms.
\smallskip
\begin{center}
\begin{tabular}{|r|ll|ll|}
\hline
    $L$ & $\mbox{Re}(\kappa_{1})$ & $\mbox{Im}(\kappa_{1})$
& $\mbox{Re}(\kappa_{2})$ & $\mbox{Im}(\kappa_{2})$ \cr
\hline
 8 & 0.152156(10)  & 0.004046(10)   & 0.154195(104) & 0.006085(127) \cr
12 & 0.150802(7)   & 0.001733(10)   & 0.151652(86)  & 0.002615(137) \cr
16 & 0.150322(7)   & 0.000948(5)    & 0.150913(38)  & 0.001352(41) \cr
20 & 0.150095(5)   & 0.000595(7)    & 0.150397(55)  & 0.000875(36) \cr
24 & 0.149972(3)   & 0.000414(5)    & 0.150198(40)  & 0.000574(48) \cr
\hline
\end{tabular}
\end{center}

\newpage
{\noindent \bf Table 3:}
The angle $\varphi_{1,2}$ between the first and second Fisher zeroes as
obtained from the jackknifed multi-histograms. Mean field theory
predicts $\varphi = \pi /4 \simeq 0.785$.
\smallskip
\begin{center}
\begin{tabular}{|r|l|}
\hline
$L$ & $ \varphi_{1,2}$  \cr
\hline
 8 &  $ 0.785(62) $\cr
12 &  $ 0.804(139) $\cr
16 &  $ 0.599(89) $\cr
20 &  $ 0.747(175) $\cr
24 &  $ 0.617(246) $\cr
\hline
mean & $ 0.731(45) $ \cr
\hline
\end{tabular}
\end{center}

\vspace{0.5cm}
{\noindent \bf Table 4:}
The positions of the first two Lee--Yang zeroes as obtained from the
jackknifed multihistograms at $\kappa = 0.149703$.
The real part of the zeroes is always zero.
\smallskip
\begin{center}
\begin{tabular}{|r|l|l|}
\hline
  $L$ & $\mbox{Im}(h_{1})$ &  $\mbox{Im}(h_{2})$ \cr
\hline
8  &  0.0022294(32)   &  0.0049488(71) \cr
12 &  0.0006367(23)   &  0.0014111(22) \cr
16 &  0.0002637(9)    &  0.0005845(33) \cr
20 &  0.0001327(7)    &  0.0002949(18) \cr
24 &  0.0000749(5)    &  0.0001656(7) \cr
\hline
\end{tabular}
\end{center}

\newpage

%------------------------------------------------------------------------------
\section*{Figures}
%------------------------------------------------------------------------------

{\noindent \bf Fig. 1:~}
(a) The imaginary part of the Fisher zeroes closest to the
real $\kappa$ axis
vs. the logarithm of the lattice size $L$. The straight line is a fit to
all points with slope -2.088(6), slightly differing from the
expected value $-1/\nu = -2$. This difference is due to logaritmic
corrections to scaling as shown in (b), where we plot $\ln (L^2 {\rm Im}
\kappa_1)$ vs. $\ln \ln L$.

\vspace{12pt}
{\noindent \bf Fig. 2:~}
The impact angle $\varphi_{1,2}$ of the closest two Fisher zeroes,
compatible  with the expected value $\pi/2$ (full line) within the error
(shaded area).

\vspace{12pt}
{\noindent \bf Fig. 3:~}
(a) The imaginary part of the Lee--Yang zeroes closest to $h=0$
vs. the logarithm of the lattice size,
for $\kappa=0.149703$. The straight line is a fit to
all points with slope -3.083(4), slightly differing from the
expected value $-1/\delta = -3$. This difference is due to logaritmic
corrections to scaling as shown in (b), where we plot $\ln (L^3 {\rm
Im} h_1)$ vs. $\ln \ln L$.
\end{document}